\documentstyle[11pt]{article}

\newcommand{\eqalign}[1]{
\null \,\vcenter {\openup \jot \ialign {\strut \hfil
$\displaystyle {
##}$&$\displaystyle {{}##}$\hfil \crcr #1\crcr }}\,}

\newtheorem{definition}{Definition}
\newtheorem{thm-def}{Definition-Theorem}

\newcommand{\Str}{\rm Str\,}
\newcommand{\s}{\sigma}
\renewcommand{\t}{\tau}
\newcommand{\p}{product integral}

\newcommand{\beq}{\begin{equation}}
\newcommand{\eeq}{\end{equation}}
\newcommand{\be}{\begin{equation}}
\newcommand{\ee}{\end{equation}}  

\title{\nopagebreak
\begin{flushright}
\tenrm UCTP101.00
\end{flushright}\vskip0.3in
\nopagebreak
\large \bf  Supersymmetric Wilson Lines and Loops, \\ and \\
Super Non-Abelian Stokes Theorem}
\author{Robert L. Karp\thanks{e-mail address:
karp@physics.uc.edu},\,
Freydoon Mansouri\thanks{e-mail address: mansouri@uc.edu}\\
{\it \small
Physics Department, University of
Cincinnati, Cincinnati, OH 45221}\\
}
\date{}

\begin{document}

\maketitle

\begin{abstract}
We generalize the standard product integral formalism to
incorporate Grassmann valued matrices and show that the resulting
supersymmetric product integrals provide a natural framework for
describing supersymmetric Wilson lines and Wilson loops. We
use this
formalism to establish the supersymmetric version of the
non-Abelian  
Stokes theorem.
\end{abstract}

\section{Introduction}

The notion of Wilson loop~\cite{rone,rtwo} provides a systematic method of
obtaining gauge invariant observables. Its standard applications range
from particle phenomenology and lattice field theories to strings and
topological gauge theories. More recently, in the context of the AdS/CFT
correspondence~\cite{mal1}, an interesting connection between Wilson loops
in supersymmetric gauge theories and membranes in supergravity theories
has been suggested~\cite{mal}. In view of this and other important
developments in supersymmetric gauge theories, it is natural to ask
whether the notions of Wilson line and Wilson loop permit a supersymmetric
generalization. Some formal work in this direction was carried out early
in the development of supersymmetric gauge theories~\cite{e1,e2,e3,e4,e5}.
There are also some recent suggestions in the $N=4$ case \cite{gross}. In
contrast to these attempts, our aim is to construct supersymmetric Wilson
lines and Wilson loops in terms of supersymmetric product integrals. For
non-supersymmetric gauge theories, it has been shown recently~\cite{kmr}
that standard product integrals~\cite{dollard} provide a natural framework
for describing Wilson lines and Wilson loops. This is because they have a
built-in feature for keeping track of the order of matrices in path
ordered quantities. The main purpose of the present work is to extend
these results to theories which involve supersymmetric matrices. Thus, our
construction of supersymmetric Wilson lines and loops is the natural
supersymmetric extension of the definition of their non-supersymmetric
counter parts. This will permit us to give, among other things, an
unambiguous proof of the supersymmetric version of the non-Abelian Stokes
theorem.

To provide a supersymmetric generalization of the notions of Wilson line
and Wilson loop in terms of product integrals, we must address a number of
questions. The first among these has to do with the fact that in
supersymmetric gauge theories, the superfields have values in a Grassmann
algebra. To be able to explore the properties of these theories in terms
of product integrals, we must first ensure that Grassmann valued product
integrals exist. We address this question in Section 2, where we construct
supersymmetric product integrals and explore their properties. In Section
3, we use this formalism to define supersymmetric Wilson lines and Wilson
loops. In Section 4, we construct a surface integral representation for
the supersymmetric Wilson loop, thus establishing the supersymmetric
version of the non-Abelian Stokes theorem. As a further confirmation of
this theorem, in Section 5, we show the gauge covariance of the surface
integral representation of the super Wilson loop operator. Section 6 is
devoted to concluding remarks.

\section{Supersymmetric Product Integrals}

Comprehensive accounts of ordinary product integrals and their
applications exist in the literature~\cite{kmr,dollard}.
Here we mention in passing that the justification
for the word ``product'' lies in the property that the \p\ is to
the product
what the
ordinary integral is to the sum and that one of their most common
applications is in solving systems of linear
differential equations of the form
\beq\label{e1} 
{\bf y'(s)} = A(s)\,{\bf y(x)},\quad {\bf y(s_0)} = {\bf y_0}. 
\eeq 
The solution of this system can be constructed in terms of the
limit of the finite
ordered product \cite{dollard}: $
\Pi_p(A)=\prod_{k=1}^{n}e^{A(s_k)\Delta s_k}$. In this expression,
$\Delta s_k = s_k -s_{k - 1}$ for $k = 1,...,n$, where $\{ s_0, s_1,
...., s_n \}$ is a partition of the real interval $[a,b]$.
In the limit of large $n$ and under suitable 
conditions, this ordered product
leads to the definition of the product integral.

The properties of standard product integrals rest heavily on the Banach
algebra structure of matrix valued functions~\cite{dollard}. In
supersymmetric theories, the corresponding matrices take values in a
Grassmann algebra. Since product integrals are products of the
exponentials of the matrix valued functions, and in a supersymmetric
theory the exponents must necessarily belong to the even part of the
Grassmann algebra, we expect intuitively that all the properties of
standard product integrals can be extended to supersymmetric product
integrals. To put this on firm mathematical foundation, we must specify a
suitable norm on the Grassmann algebra, with respect to which
supersymmetric matrices also acquire a Banach algebra structure.

The Banach algebra structure of the Grassmann algebra is well
known~\cite{freund}. Consider for definiteness the finite dimensional
Grassmann algebra generated by the anticommuting quantities $\theta^1,
\theta^2,\ldots, \theta^p$. In this case, a generic element of the algebra
can be written as a linear combination of the products $\theta^{i_1}
\theta^{i_2}\ldots \theta^{i_k}$, $k=0,\ldots,p$, with complex
coefficients $a_{i_1i_2\ldots i_k}$. As a complex vector space the
Grassmann algebra is $2^p$ dimensional. A norm on the above vector space
(more precisely a valuation of the algebra) can be defined as the sum of
the moduli of the coefficients. For example in the Grassmann algebra
generated by a single element $\theta$, the norm of the generic element
$x=a+b\theta$, $a,b\in{\bf C}$, is $||x||=|a|+|b|$, with $|a|,|b|$ the
complex moduli.  From this definition, one can show that the norm of the
product of any two elements $x$ and $y$ of the Grassmann algebra satisfies
the inequality: $||x\cdot y||\leq||x||\cdot||y||$. This result is true not
only for the above simple example but for the general Grassmann algebra
generated by $\theta^1, \theta^2,\ldots, \theta^p$.  It is also
straightforward to show that this norm is complete. In other words, with
respect to this norm, the Grassmann algebra becomes a Banach algebra. As
we will see below, this allows us to extend to supersymmetric product
integrals most of the theorems which apply to ordinary product
integrals~\cite{dollard}.

Having specified a suitable norm on the Grassmann algebra, we
turn to the
construction of supersymmetric product
integrals and to the study of some of their basic properties.
\begin{definition}
Let $\Gamma:[a,b]\rightarrow {\bf C}^{1|p}_{n\times n}$ be an
$n\times
n$
matrix valued function with entries in the complex superspace ${\bf
C}^{1|p}$. Let $P=\{s_0,s_1,\ldots,s_n\}$ be a partition of the
interval
$[a,b]$, with $\Delta s_k=s_k-s_{k-1}$ for all $k=1,\ldots,n$.
\begin{description}
\item[\it (i)] $\Gamma$ is called a {\em step function} iff there
is a partition
$P$ such that $\Gamma$ is constant on each open subinterval
$(s_{k-1},s_k)$, for
all $k=1,\ldots,n$.
\item[\it (ii)]The {\em point value approximant } $\Gamma_P$
corresponding to
the function $\Gamma$ and partition P is the step function taking
the
value
$\Gamma(s_k)$ on the interval $(s_{k-1},s_k]$ for all
$k=1,\ldots,n$.
\item[\it (iii)]If $\Gamma$ is a step function, then we define
the function
$E_{\Gamma}:[a,b]\rightarrow {\bf C}^{1|p}_{n\times n}$ by
$E_{\Gamma}(x):=
e^{\Gamma(s_k)(x-s_{k-1})} \ldots e^{\Gamma(s_2)\Delta s_2}
e^{\Gamma(s_1)\Delta
s_1}$
for any $x\in (s_{k-1},s_k]$, for all $k=1,\ldots,n$, and
$E_{\Gamma}(a):=I$.
\end{description}
\end{definition}

Based on the product integral formalism developed for ordinary matrices
\cite{dollard},
we want the functions $E_{\Gamma}$ to converge to the product integral as
the partition of $[a,b]$ is refined. For the proof of the existence of the
supersymmetric product integral we need some preliminary results. We start
with estimating the norm of $E_{\Gamma}$. This requires one more
ingredient, namely the norm of a Grassmann algebra valued matrix. This
will be defined in analogy with that of ordinary matrices: for any
${n\times n}$ matrix $\Gamma$ as above, we define
\be
||\Gamma||_M=\sup_{x\in {\bf C}^{n|p}\,\, ||x||_n\leq 1}
\frac{||\Gamma\,x||_n}{||x||_n},
\ee
where $||x||_n$ refers to the norm of $x$ as an element in $({\bf
C}^{1|p})^n\equiv {\bf C}^{n|p}$. For $x=(x_1,\ldots, x_n)\in {\bf
C}^{n|p}$ we
can define
$||x||_n$ for example by $||x||_n=\sum_{i=1}^n||x_i||$. Some
clarifications are necessary at this point. The Grassmann algebra ${\bf
C}^{1|p}$ is a ${\bf C}$ vector space, but it is not a field. As a result
${\bf C}^{n|p}$ is not a vector space over ${\bf C}^{1|p}$, but only a
rank $n$ module. Though it is a ${\bf C}$ vector space, it has no
canonical norm on it. With all these preparations we have:
\be\eqalign{
||E_{\Gamma}(x)||_M&=||e^{\Gamma(s_k)(x-s_{k-1})} \ldots
e^{\Gamma(s_2)\Delta s_2}
e^{\Gamma(s_1)\Delta s_1}||_M\leq \cr
&\leq||e^{\Gamma(s_k)(x-s_{k-1})}||_M \ldots
||e^{\Gamma(s_1)\Delta s_1}||_M\leq e^{\int_a^x ||\Gamma(s)||_M d s}.
}
\ee
In summary, we have obtained the following result:
\be\label{est}
||E_{\Gamma}(x)||_M\leq e^{\int_a^x\! d s  ||\Gamma(s)||_M}.
\ee
As a final preparation, we prove the following lemma:
Let $\Gamma_1, \Gamma_2 :[a,b]\rightarrow {\bf C}^{1|p}_{n\times
n}$ be
step-functions.
Then,
\be\label{pi}
E_{\Gamma_1}(x)-E_{\Gamma_2}(x)=E_{\Gamma_2}(x)\int_a^x \! d s\,
E_{\Gamma_2}^{-1}(s)[\Gamma_1(s)- \Gamma_2(s)]E_{\Gamma_1}(s).
\ee
To prove this, we define
$G(x)=E_{\Gamma_2}^{-1}(x)E_{\Gamma_1}(x)$. It follows
immediately
that
$G(a)=I$ and $G(x)$ is continuous, and differentiable except for
the
division points of the partitions associated to $\Gamma_1$ and
$\Gamma_2$. As a
result, except for the division points, we have 
\be
G'(x)=E_{\Gamma_2}^{-1}(x)[\Gamma_1(x)-
\Gamma_2(x)]E_{\Gamma_1}(x).
\ee
The quantity $G(x)$ is continuous and is continuously
differentiable on each open
division subinterval. Then, using the fundamental theorem of
calculus
on the
subintervals and piecing the results together, we get:
\be
G(x)=I+\int_a^x \! d s\, E_{\Gamma_2}^{-1}(s)[\Gamma_1 (s)-
\Gamma_2 (s)]E_{\Gamma_1}(s).
\ee
Multiplication from the left by $E_{\Gamma_2}(x)$ leads to Eq.
(\ref{pi}).

We are now in a position to define the supersymmetric product integral,
and prove its existence:

\begin{thm-def}
Given a continuous function $\Gamma:[a,b]\rightarrow {\bf
C}^{1|p}_{n\times
n}$ and a sequence of step functions
$\{\Gamma_n\}$, which converges to $\Gamma$ in the
$L^1([a,b])$ sense, then the sequence $\{E_{\Gamma_{n}}(x)\}$
converges
uniformly on $[a,b]$ to a matrix called the {\em supersymmetric
\p }\ of $\Gamma$ over
$[a,b]$.
\end{thm-def}

To prove the existence of super product integrals, we must
demonstrate the convergence of the sequence
$\{E_{\Gamma_{n}}(x)\}$.
By the lemma given above, we have
\be
E_{\Gamma_n}(x)-E_{\Gamma_m}(x)=E_{\Gamma_m}^{-1}(x)\int_a^x \! d
s\,
E_{\Gamma_m}(s)[\Gamma_n(s) -
\Gamma_m(s)]E_{\Gamma_n}(s).
\ee
We can estimate the norm of the left-hand-side (lhs) as follows:
\be 
||E_{\Gamma_n}(x)-E_{\Gamma_m}(x)||_M\leq
||E_{\Gamma_m}^{-1}(x)||_M\int_a^x \! d s\,
||E_{\Gamma_m}(s)||_M\,||\Gamma_n(s) -
\Gamma_m(s)||_M\,||E_{\Gamma_n}(s)||_M .
\ee
Using Eq. (\ref{est}), we can estimate the difference of the
norms   as
\be
||E_{\Gamma_n}(x)-E_{\Gamma_m}(x)||_M\leq e^{2\int_a^b \! d s
||\Gamma_m(s)||_M}e^{\int_a^b
\! d s ||\Gamma_n(s)||_M}\int_a^x \! d s\, ||\Gamma_n(s) -
\Gamma_m(s)||_M.
\ee
Since $\{\Gamma_n\}$ converges to $\Gamma$ in the $L^1([a,b])$
sense, the
first two
terms on the rhs are bounded, and the sequence $\{\Gamma_n\}$ is
Cauchy in the
$L^1([a,b])$ sense. Accordingly, the rhs goes to zero as
$m,n\rightarrow
\infty$. Since the rhs is independent of $x$, $\{E_{\Gamma_n}\}$
is uniformly
Cauchy, hence uniformly convergent. This establishes the 
existence the supersymmetric product integral. 
To prove the uniqueness of the limit, we estimate the difference
$||E_{B_n}(x)-E_{C_n}(x)||_M$ for two sequences $\{B_n\}$ and $\{C_n\}$,
converging to $\Gamma$ in the $L^1$ sense. Proceeding as we did above, it
is
immediate that $\{E_{B_n}\}$ and $\{E_{C_n}\}$ have the same limit. This
concludes the proof of the existence and uniqueness of the supersymmetric
product integral.

The structure of the supersymmetric product integrals described
above
permits the generalization of some of the well-known theorems of
product
integration \cite{dollard} to the supersymmetric case. Here we give
a summary
of the results which are relevant to the proof of the
supersymmetric
non-Abelian Stokes' theorem. The proofs and further discussion of
these results will be given elsewhere~\cite{km2}.

We will follow the notation and
the conventions of \cite{kmr} as much as possible. Let
$\Gamma:[a,b]\rightarrow {\bf
C}^{1|p}_{n\times n}$ be a continuous Grassmann valued function.
For
any $x\in [a,b]$, we express the supersymmetric product integral from
$a$ to
$x$ as
\beq
F(x,a):=\prod_{a}^{x} e^{\Gamma(s)d s}. 
\eeq 
Then, $F$ satisfies
the integral equation: 
\beq F(x,a)=1+\int_{a}^{x}\,ds\,\Gamma(s)F(s,a) \label{ei}.
\eeq 
It is also a solution of the initial
value problem: 
\beq \label{bv}
\frac{dF}{dx}(x,a)=\Gamma(x)F(x,a),
\quad F(a,a)=I.
\label{eii} 
\eeq 
The determinant of a supersymmetric product integral is
given by 
\beq 
\det\left(\prod_{a}^{x} e^{ \Gamma(s)ds}\right)=e^{
\int_{a}^{x}\, \Str \Gamma(s)ds}, 
\eeq  
where Str stands for
supertrace. 
The intuitive 
composition rule holds:
\beq\label{cr} \prod_{a}^{b} e^{
\Gamma(s)ds}=\prod_{c}^{b} e^{ \Gamma(s)ds}\prod_{a}^{c}
e^{\Gamma(s)ds}. 
\eeq
It is possible to 
differentiate with
respect to the endpoints:
\beq\label{thm2}
\frac{\partial}{\partial x}\left(\prod_{y}^{x} e^{
\Gamma(s)ds}\right) =
\Gamma(x)\prod_{y}^{x} e^{ \Gamma(s)ds},\quad\quad
\frac{\partial}{\partial y}\left(\prod_{y}^{x} e^{
\Gamma(s)ds}\right) =
-\prod_{y}^{x} e^{ \Gamma(s)ds} \Gamma(y). 
\eeq
The
L-derivative of ordinary product
integrals~\cite{dollard} can be extended to super product
integrals: for  a non-singular differentiable Grassmann valued
function  
$\Gamma:[a,b]\rightarrow {\bf C}^{1|p}_{n\times n}$, we define
\beq\label{lo}
L\, \Gamma(x):=\Gamma'(x)\Gamma^{-1}(x),
\eeq
where prime indicates differentiation with respect to $x$.
Defining 
\beq
P(x)=\prod_{a}^{x} e^{ \Gamma(s)ds},
\eeq
and using Eq. (\ref{bv}), we can extend the analog of the fundamental
theorem of
calculus to
super product integrals:
\beq\label{thm3}
\prod_{a}^{x} e^{ (LP)(s)ds}=P(x)P^{-1}(a).
\eeq
The proof of the super non-Abelian Stokes' theorem given below
will rely heavily on the contents of the next three theorems.
The first one is the {\em sum rule}. With $P(x)=\prod_{a}^{x} e^{
\Gamma_1(s)ds}$, we have
\beq\label{sr}
\quad \prod_{a}^{x} e^{\,
[\Gamma_1(s)+ \Gamma_2(s)]ds}=P(x)\prod_{a}^{x}
e^{\,P^{-1}(s)\Gamma_2(s)P(s)ds}.
\eeq
The second one is the {\em similarity rule}:
\beq\label{si}
\quad P(x)\left( \prod_{a}^{x} e^{
\Gamma_2(s)ds}\right)P^{-1}(a)=\prod_{a}^{x}
e^{\, [LP(s)+P(s)\Gamma_2(s)P^{-1}(s)]ds}.
\eeq
Finally, the third one is {\em differentiation with respect to a
parameter}. Given a Grassmann valued  function
$\Gamma:[a,b]\times[c,d]\rightarrow {\bf C}^{1|p}_{n\times n}$
satisfying
proper
differentiability conditions, and given
$P(x,y;\lambda)=\prod_{y}^{x} e^{
\Gamma(s;\lambda)ds}$, we have:
\beq\label{dp}
\frac{\partial}{\partial \lambda} P(x,y;\lambda)=\int_{y}^{x}
d s\,P(x,s;\lambda)\frac{\partial \Gamma}{\partial
\lambda}(s;\lambda)P(s,y;\lambda).
\eeq
\label{thmp}

\section{Supersymmetric Wilson Lines and Loops}

Our results for supersymmetric product integrals are fairly general. In
this section, we will use them as a basis to provide a natural and
mathematically sound definition of supersymmetric Wilson lines and loops.
To this end, we introduce our notations in a manner which naturally arises
in supersymmetric gauge theories. We focus on the supersymmetric Wilson
loop first. Consider an oriented manifold $M$ and a closed path $C$ in
$M$. For simplicity, we assume that the target space is a simply connected
manifold $M$, i.e. $\pi_1(M)=0$. This insures that the loop may be taken
to be the boundary of an orientable two dimensional surface $\Sigma$ in
$M$. It will be convenient to describe the properties of such a 2-surface
in terms of local coordinates $\s^0=\t$ and $\s^1=\s$. So, for the points
of the manifold $M$, which lie on $\Sigma$, we have $x = x(\s, \t)$.

Let, in standard two component spinor notation~\cite{wess}, the local
coordinates of a superspace be given by
$z^M=(x^{\alpha \dot{\alpha}}, \theta^{\alpha},
\theta^{\dot{\alpha}})$. Also let the
components of a supersymmetric connection $\Gamma$ be given by $\Gamma_M$.
In terms of local coordinates, the connection $\Gamma$ is a Lie
superalgebra valued superform, which can be expressed as $\Gamma =
dz^M\Gamma_M$. From the point of view of covariance under supersymmetry
transformations, it is more convenient to express $\Gamma$ in a basis in
which the exterior derivative operator $d=dz^M\,\partial_M$ maps
superfields to superfields~\cite{wess}. So, we shall work, instead, in the
basis where $d=e^A D_A$, with $D_A$ the supersymmetric covariant
derivative, and $e^{A}(z)=dz^Me_M^{\quad\!\!\! A}(z)$. In this expression,
$e_M^{\quad\!\!\! A}(z)$ are the well-known super-beins. Thus, we have
\beq
\Gamma(z) = dz^M\Gamma_M(z)=e^{A}(z)\Gamma_A(z).
\eeq
To describe Wilson lines and Wilson loops, we need
the pull-back of this quantity on the path $C$ in $M$,
described by an intrinsic parameter $s$:  $x^{\alpha \dot{\alpha}} =
x^{\alpha \dot{\alpha}}(s)$,
$\theta^{\alpha}=
\theta^{\alpha}(s)$, and $\theta^{\dot{\alpha}}=\theta^{\dot{\alpha}}(s)$.
In terms of the embedding map
$i:C\rightarrow
M$ we have:
\beq\label{dd}
\Gamma(s)=i^*\Gamma(z)=\partial_{s} z^M(s)\,\Gamma_M(z(s)).
\eeq
Similarly, to obtain the pull-back of $\Gamma$ on the 
2-surface, we use the supersymmetric vielbeins:
\beq
\Gamma_{a}=v_{a}^{A}\Gamma_A;\quad\quad v_{a}^{M}=\partial_a z^M;
\quad
v_{a}^{A}=v_{a}^{M}e_M^{\quad\!\!\!
A}(z).
\eeq
It is the quantity $\Gamma=\Gamma(s)ds$ or 
$\Gamma=\Gamma_{a}d\s^a$ that we will
identify with the matrix valued functions of the supersymmetric
product integral formalism described above. 
The corresponding pull-backs of the components of the
supersymmetric covariant derivative on the line and on the 2-
surface are given, respectively, by $\frac{\partial}{\partial s}$
and
\beq
D_a = v^A_a\,D_A = \frac{\partial}{\partial\sigma^a}= \partial_a.
\eeq

The components of the supersymmetric field strength $F_{ab}$ on
the
2-surface can be computed in two different ways. The first method
is the obvious pull-back of
the
target space supersymmetric field strength:
\beq \label{g1}
F_{ab} = v_{a}^{A}v_{b}^{B}F_{BA}=v_{a}^{M}v_{b}^{N}F_{NM}. 
\eeq
The second method is to make use of the pulled-back connection 
$\Gamma_{a}$ given above:
\beq\label{g2}
F_{ab} =
\partial_a\Gamma_b-\partial_b\Gamma_a + [\Gamma_a,\Gamma_b].
\eeq
To show the consistency of the above two expresions, 
multiply both (\ref{g1}) and
(\ref{g2}) with the wedge product of differential forms ${1\over
2}d\s^a\wedge d\s^b$ to get the corresponding field strength
two-forms on the two-surface. Then, the consistecy amounts to showing that the
two field strength expressions are equal. Since on the two-surface
$d\s^av_{a}^{M}=dz^{M}$, Eq. (\ref{g1}) becomes ${1\over
2}dz^{M}dz^{N}F_{NM}$. Moreover, $d\s^a\partial_a=d$ on the two-surface,
so that Eq. (\ref{g2}) becomes $d\Gamma-\Gamma^2$. But
this expression is equal to the previous one by definition \cite{wess}.

Consider now the continuous map
$\Gamma :[a,b]\rightarrow {\bf R}^{1|4}_{n\times n}$, where the
latter is an $n$ by
$n$
matrix valued function, with entries in the superspace ${\bf
R}^{1|4}$, 
corresponding to the pull-back on the path C.
Then, we define the
supersymmetric Wilson line in terms of a super product integral
as follows:
\beq
{\cal P} e^{\int_{a}^{b} \Gamma(s)ds}
\equiv \prod_{a}^{b} e^{\Gamma(s)ds},
\eeq
where $\cal P$ indicates path ordering as defined by the super
product integral on the right-hand-side.
Anticipating that we will identify the closed path $C$ over which
the
Wilson loop is defined with the boundary of a 2-surface, it is
convenient
to work from the beginning with Wilson lines depending on a
parameter.
Define $\Gamma_a:[\s_0,\s_1]\times[\t_0,\t_1]\rightarrow
{\bf
R}^{1|4}_{n\times n}$, where $[\s_0,\s_1]$ and $[\t_0,\t_1]$ are
the range
of the local coordinates on the two surface $\Sigma$.
For later convenience, we also define the following elementary
supersymmetric Wilson lines:
\beq\label{dps}
{P}(\sigma,\s_0;\t)= \prod_{\s_0}^{\s} e^{
v_1^A\Gamma_A(\sigma';\t)d\sigma'},\qquad
{ Q}(\sigma;\t,\t_0) = \prod_{\t_0}^{\t} e^{
v_0^A\Gamma_A(\sigma;\t')d\t'}.
\label{dqs}
\eeq

To prove the supersymmetric version of the non-Abelian Stokes
theorem, we
want to make use of super product integration techniques to
express the
super Wilson loop operator as an integral over a two dimensional
surface bounded by the corresponding loop. 
For this purpose, we define the super Wilson
loop operator as
\beq
W_s[C] = {\cal P} \exp{(\oint_C \Gamma(\t) d\t)}\equiv e^{\oint_C
i^*(dz^M\,\Gamma_M)}.
\label{s24}
\eeq
In this expression, as in Eq. (24), 
$i^*$ denotes the pull-back of the embedding $i:C\rightarrow M$.
We have written this expression in a notation familiar from the physics
literature. It is to be understood, however, that the right-hand-side is
to be composed of the super product integrals as given in Eq. (29) above.
The expression for the supersymmetric Wilson loop depends on the homotopy
class of the loop $C$ in $M$. We can, therefore, parameterize $C$ in any
convenient manner consistent with its homotopy class. In particular, we
can break up the closed path into piecewise continuous segments, along
which either $\s$ or $\t$ remains constant. The composition rule for super
product integrals given by Eq. (\ref{cr}) ensures that this break up of
the super Wilson loop into super Wilson lines does not depend on the
intermediate points chosen on the closed path. Inspired by the typical
paths which are used in the actual computations of of both ordinary and 
supersymmetric
Wilson loops (see e.g. \cite{mal,kmr}), we break up the super Wilson
loop into a product of four
super Wilson lines. Using the same notation as in the non-supersymmetric
case \cite{kmr}, we write
\beq
W_s[C] = W_4\, W_3\, W_2\, W_1 .
\label{w}
\eeq
In this expression, $W_k$, $k=1,..,4$, are super Wilson lines
such that
$\t = const.$ along $W_1$ and $W_3$, and $\s = const.$ along
$W_2$ and $W_4$. We emphasize that $\s = const.$ and $\t =
const.$ are arbitrary curves.

To see the advantage of parameterizing the closed path in this
manner,
consider the  exponent of Eq. (\ref{s24}).
Along each segment, only one 
of the terms is non-vanishing.
For example, along the segment $[\s_0 ,\s]$,
we have $\t ' = \t_0 = const$.
As a result, we obtain:
\beq
W_1 =P(\s,\s_0; \t_0),\quad
W_2 =Q(\s; \t,\t_0),\quad
W_3 = P^{-1}(\s,\s_0; \t),\quad
W_4 = Q^{-1}(\s_0; \t,\t_0).
\label{s29}
\eeq
Using these expressions, the supersymmetric Wilson loop can be
expressed as
\beq
W_s[C]={ Q}(\sigma_0;\t,\t_0)^{-1}{ P}(\sigma,\s_0;\t)^{-1}
{ Q}(\sigma;\t,\t_0){ P}(\sigma,\s_0;\t).
\label{sw}
\eeq 

For definiteness, in the rest of the paper we will confine ourselves to
the case in which the 2-surface, $\Sigma$, can be covered by a single
coordinate patch.  If $\Sigma$ requires more than one patch to be covered,
then using partition of unity and the product integral composition rule,
Eq. (\ref{cr}), it is straightforward to extend our upcoming reasonings.

\section{Super Non-Abelian Stokes Theorem}

As an application of the supersymmetric \p\ formalism, we prove
the supersymmetric version of the non-Abelian Stokes theorem~\cite{km1}.  
The proof makes essential use of the generalized theorems listed in the
previous paragraphs, and is the supersymmetric version of one of the
proofs given for the non-supersymmetric case in reference~\cite{kmr}. The
other proof give in this  reference can also be extended to the
supersymmetric case and will be given in a subsequent work~\cite{km2}.

We start with the form of $W_s[C]$ given in Eq. (\ref{sw}) and take
its
derivatives with respect to the parameter $\t$:
\begin{eqnarray}
{\partial W_s[C] \over {\partial \t}}&=&\partial_\t
Q^{-1}(\sigma_0;\t,\t_0)
P^{-1}(\s,\s_0;\t) Q(\s;\t,\t_0) P(\s,\s_0;\t_0)+
\nonumber\\
&&+Q^{-1}(\sigma_0;\t,\t_0)
\partial_\t P^{-1}(\s,\s_0;\t) Q(\s;\t,\t_0) P(\s,\s_0;\t_0)+
\nonumber\\
&& +
Q^{-1}(\sigma_0;\t,\t_0)P^{-1}(\s,\s_0;\t) \partial_\t
Q(\s;\t,\t_0)
P(\s,\s_0;\t_0).
\end{eqnarray}
Here, we have made use of the fact that $P(\s,\s_0;\t_0)$ is
independent
of $\t$. As a preparation for using Eq. (\ref{thm3}), we
start with Eq. (\ref{lo}) for $W_s[C]$, and make use of Eq. (\ref{thm2})
to get
\begin{eqnarray}
L_\t W_s[C]={\partial W_s [C]\over {\partial \t}}
W_s[C]^{-1}=\!&T^{-1}(\s;\t)\,[\Gamma_0(\s;\t)-
P(\s,\s_0;\t)\Gamma_0(\s_0;\t)P^{-1}(\s,\s_0;\t)-
\nonumber\\
&-\partial_\t
P(\s,\s_0;\t)P^{-1}(\s,\s_0;\t)]\,T(\s;\t),
\label{n1}
\end{eqnarray}
In this expression, $T(\s; \t) = P(\s,\s_0; \t)\, Q(\s_0;
\t,\t_0)$.
Next, by means of differentiation with respect to a parameter
given by Eq. (\ref{dp}), we evaluate the derivative of
the super \p\ $P(\s, \s_0; \t)$
with respect to the parameter $\t$:
\beq
\partial_\t P(\s,\s_0;\t)=\int_{\s_0}^\s
d\s'P(\s,\s';\t)\partial_\t \Gamma_1
(\s';\t)P(\s',\s_0;\t).
\eeq
Then, after some simple manipulations using the defining
equations
for the various terms in Eq. (\ref{n1}), we get:
\beq
T^{-1}(\s;\t)\partial_\t
P(\t)P^{-1}(\t)T(\s;\t)=\int_{\s_0}^\s d\s'
T^{-1}(\s';\t)\partial_\t \Gamma_1
(\s';\t)T(\s';\t).
\label{n2}
\eeq
Using Eq.  (\ref{thm2}) and the fact that $P(\s_0,\s_0;\t)=1$,
we can rewrite
the rest of
Eq. (\ref{n1}) also as an integral:
\begin{eqnarray}\label{n3}
&T^{-1}(\s;\t)[\Gamma_0(\s;\t)-
P(\s,\s_0;\t)\Gamma_0(\s_0;\t)P^{-1}(\s,\s_0;\t)]T(\s;\t)=
\nonumber\\
&=\int_{\s_0}^\s d\s'\, P^{-1}(\s',\s_0;\t)(\partial_\t
\Gamma_0(\s',\t)
+[\Gamma_0(\s',\t),\Gamma_1(\s',\t)])P(\s',\s_0;\t).
\end{eqnarray}
Combining Eqs. (\ref{n1}), (\ref{n2}), and (\ref{n3}), we obtain:
\beq
L_\t W_s[C]=\int_{\s_0}^\s d\s'
T^{-1}(\s',\t)F_{01}(\s',\t)T(\s',\t).
\eeq
Here $F_{01}$ is the field strength component as defined in
Eq. (\ref{g2}),
but based on the discussion in that paragraph, we know that it
also equals
the pull-back of the supersymmetric field strength to the
surface. 
Using Eq. (\ref{thm3}), we are immediately led to the 
supersymmetric version of the non-Abelian
Stokes theorem:
\beq\label{nas}
W_s[C]=\prod_{\t_0}^{\t} e^{\int_{\s_0}^{\s} T^{-1}(\s';\t')
F_{0 1}
(\s';\t') T(\s';\t')d\s' d\t'}.
\eeq
Recalling the antisymmetry of the components of the field
strength, we can rewrite this expression in a more familiar
reparameterization invariant form
\beq
W_s[C]={\cal P_\t}e^{\oint \Gamma}=\prod_{\t_0}^{\t}    
e^{ {1\over 2}\int_\Sigma d\s^{ab}\,T^{-1}(\s;\t)F_{ab}
(\s;\t) T(\s;\t)},
\eeq
where $d\s^{ab}$
is the area element of the 2-surface.
Despite appearances, it must be remembered that $\s$ and $\t$
play very different roles in this expression. 

The above result also applies to the special case in which the
gauge group is Abelian. In that case, however, since the
corresponding
matrices commute, the machinery of the super product integrals is
not needed, and one can establish the super Stokes theorem
directly~\cite{am}.

\section{Gauge Covariance of the Super Loop Operator}

To demonstrate the gauge covariance of the supersymmetric Wilson loop
operator and its 2-surface representation, we must show how the
supersymmetric Wilson line transforms under gauge transformations. For
this, we need to know, in turn, how the pull-back of the connection
$\Gamma_A(z)$ transforms.
The transformation
properties of the connection itself follows from that of the
vector superfield~\cite{wess}: 
$e^{V'}=e^{-i\Lambda^\dagger}e^V e^{\Lambda}.$
More specifically, we have 
\beq
{\Gamma '}(z)={ g(z)}\Gamma(z){ g(z)}^{-1}-{ g(z)}d{
g(z)}^{-1},
\eeq
where, ${g}(z)= e^{i \Lambda(z)}$ and $d=e^A(z)D_A$. As we have seen, the
pull-back of this quantity on the line is given by
$d=ds\partial_s$. Thus, we get for the transformation of the
supersymmetric connection on the line:
\beq
\Gamma '(s)={ g}(s)\Gamma(s){ g}^{-1}(s)-{ g}(s)\partial_s{
g}^{-1}(s).
\eeq
This is formally identical to that for
the plain
Yang-Mills theory~\cite{kmr}.
As a result, under a gauge transformation we obtain:
\beq
\prod_{a}^{b} e^{ ds\,\Gamma (s) }
\longrightarrow
\prod_{a}^{b}
e^{[g(s)\Gamma(s)g^{-1}(s)-g(s)\partial_s 
g^{-1}(s)]\,ds}.
\eeq
By Eq. (\ref{lo}), we have ${ g}(s)\partial_s{ g}^{-1}(s)=-L_s
g(s).$
Thus, for the gauge transformed super Wilson line we have
\beq
\prod_{a}^{b}
e^{[g(s)\Gamma(s)g^{-1}(s)+L_s g(s)]\,ds}.
\eeq
Moreover, using Eq. (\ref{si}) and recalling from Eq.
(\ref{thm3}) 
that $\prod_{a}^{b}e^{L_s g(s)\,ds}=g(b)g^{-1}(a)$, 
the gauge transformed expression takes the form
\beq
g(b)g^{-1}(a)\prod_{a}^{b} 
e^{g(a)\Gamma(s)g^{-1}(a)}.
\eeq
Finally, using the same argument as in reference~\cite{kmr},
the constant terms in the exponents can be factored out from the
 super \p . Thus, we get for the gauge transformed super Wilson
line
\beq
\prod_{a}^{b} e^{ ds\,\Gamma (s) }
\longrightarrow
g(b)\left(\prod_{a}^{b}e^{ ds\,\Gamma (s) }\right)g^{-1}(a).
\label{factor}
\eeq
We can use this result to determine the gauge transforms of
operators which are products of simple super Wilson lines.
Consider, e.g., the operator $T(\s; \t)$ which is the product of
two super Wilson lines. Applying Eq.
(\ref{factor}) to each factor, we obtain:
\beq
T(\s; \t)\longrightarrow
g(\sigma;\t)T(\s; \t) g^{-1}(\sigma_0;\t_0).
\label{gt3}
\eeq
From this, we can easily obtain the transformation properties of
the super Wilson loop operator which is also a composite of super
Wilson lines. The transformation has the same form as Eq. (\ref{factor}) with
$a=b$. 

Finally, let us consider how the surface integral representation
of super Wilson loop operator given by Eq. (\ref{nas}) transforms
under gauge transformation. From knowing how each factor in the
exponent transforms, it follows that
\beq
W_s[C] \longrightarrow
\prod_{\t_0}^{\t} e^{g(\sigma_0;\t_0)\left(\int_{\s_0}^{\s}
T^{-1}(\s';\t') F_{01}(\s';\t')T(\s';\t') dt'\right)
g^{-1}(\sigma_0;\t_0)}.
\eeq
Just as for super Wilson line, the constant terms in the
exponent factorize, so that under gauge transformations the
surface integral representation of the super Wilson loop
transforms covariantly:
\beq
W_s[C] \longrightarrow
g(\sigma_0;\t_0)\prod_{\t_0}^{\t}
e^{\int_{\s_0}^{\s} T^{-1}(\s';\t')
F_{01}
(\s';\t')T(\s';\t') dt'} g^{-1}(\sigma_0;\t_0).
\eeq

\section{Concluding Remarks}

In this work, we have presented a supersymmetric generalization of
ordinary product integral formalism. Given that Wilson lines and Wilson
loops can be expressed in terms of ordinary product integrals, we have
constructed the supersymmetric extensions of these notions for
supersymmetric gauge theories in terms of supersymmetric product
integrals. These constructions are natural in the sense that the
supersymmetric representations given in this paper reduce to the ordinary
product integral representations of standard Wilson lines and Wilson
loops.

It is hoped that this formalism provides a reliable non-perturbative means
of extracting information from supersymmetric gauge theories. In this
respect, we note that the construction of the supersymmetric Wilson lines
and Wilson loops as well as the proof of the super non-Abelian Stokes
theorem given in the previous sections are independent of any specific
physical applications. To apply these concepts to supersymmetric gauge
theories, it is necessary to clarify the physical content of the operators
such as the connection and the field strength which appear in the relevant
expressions~\cite{km1,km2}. It is well known that in supersymmetric gauge
theories the superfield strength $F_{AB}$ contains more degrees of freedom
than is required by supersymmetry and gauge invariance~\cite{grimm}. As a
result, it is necessary to impose constraints on the components of the
field strength to eliminate the unphysical degrees of freedom. This means
that in the expressions for supersymmetric Wilson lines and loops,
$\Gamma$ and $F$ must be expressed in terms of unconstrained superfields,
just as in the abelian case~\cite{am}. Such a description in terms of
unconstrained superfields already exist in the
literature~\cite{e4,wess,grimm} and can be adapted to specific
applications.

\vspace{0.3in}

This work was supported in part by the Department of Energy under the
contract number DOE-FGO2-84ER40153. We are grateful to M. Awada for
valuable input at the initial stages of this work. We would also like to
thank R. Grimm, A. Kiss and R.L. Mkrtchian for helpful communications.

\end{document}